\documentclass[11pt,a4paper]{article}
\usepackage[english]{babel}
\usepackage{graphicx,array,dcolumn,amsmath,amssymb,hhline,hyperref,color,float,theorem,mdwlist,cancel,multirow,hyperref,cite}
\usepackage[left=3cm,right=3cm,top=3cm,bottom=3cm,head=1cm,foot=1cm]{geometry}

\newcommand{\brackets}[1]{\left({#1}\right)}

\begin{document}

\title{Initial energy density of $\sqrt{s}=7$ and 8 TeV p+p collisions\\ at the LHC}

\author{M\'at\'e~Csan\'ad $^{1}$, Tam\'as~Cs\"org\H{o} $^{2,3}$, Ze-Fang Jiang $^{4}$ and Chun-Bin Yang $^{4}$\\
{\small $^1$ E\"otv\"os Lor\'and University, H -- 1117 Budapest, P\'azm\'any P. s. 1/A, Hungary}\\
{\small $^2$ Wigner RCP, H -- 1525 Budapest 114, POBox 49, Hungary}\\
{\small $^3$ EKU KRC, H -- 3200 Gy\"ongy\"os, M\'atrai \'ut 36}\\
{\small $^4$ Central China Normal Universit, 152 Luoyu Road ,Wuhan, Hubei 430079, People's Republic of China}
}

\maketitle

\begin{abstract}
Results from the RHIC and LHC experiments show, that in relativistic heavy ion collisions, a new state of
matter, a strongly interacting perfect fluid is created. Accelerating, exact and explicit solutions of
relativistic hydrodynamics allow for a simple and natural description of this medium. A finite rapidity
distribution arises from these solutions, leading to an advanced estimate of the initial energy density of
high energy collisions. These solutions can be utilized to describe various aspects of proton-proton collisions,
as originally suggested by Landau. We show that an advanced estimate based on hydrodynamics yields an initial
energy density in $\sqrt{s}=7$ and 8 TeV p+p collisions at LHC on the same order as the critical energy density from lattice
QCD, and a corresponding initial temperature around the critical temperature from QCD and the Hagedorn
temperature. The multiplicity dependence of the estimated initial energy density suggests that in
high multiplicity pp collisions at the LHC, there is large enough initial energy density to create a non-hadronic
perfect fluid.
\end{abstract}

\section{Introduction}
The first years of data taking at the experiments at the Relativistic Heavy Ion Collider (RHIC)
resulted in the discovery of a new state of matter produced in $\sqrt{s_{NN}} = 200$ GeV gold-gold
collisions~\cite{Adcox:2004mh,Adams:2005dq,Back:2004je,Arsene:2004fa}, the so-called strongly
coupled Quark Gluon Plasma (sQGP), the ``perfect liquid of quarks'': it consists of deconfined quarks,
but behaves like an almost perfect fluid (contrary to the expectations that predicted a gas-like state).  

The RHIC experiments achieved the following
``milestones''~\cite{Adare:2006nq,Adare:2006ti,Adare:2008ab,Adare:2011zr,Afanasiev:2007aa,Adare:2010ux,Adare:2011tg}
among others in discovering and characterizing this new state of
matter. The opaqueness to energetic partons is because a new state of matter is present. This matter behaves
collectively and contains deconfined quarks. Its viscosity to entropy ratio is the lowest of any known substance, and
its initial temperature is at least twice the Hagedorn temperature~\cite{Hagedorn:1965st}, the upper limiting temperature
for the description of matter based exclusively on hadronic states. This Hagedorn temperature is found in a
phenomenological manner based on just the number of hadronic states as a function of mass, however, it
is important to note that lattice QCD calculations also predict a quark deconfinement transition around
the same temperature, approximately 170 MeV~\cite{Aoki:2006br,Borsanyi:2010cj,Bazavov:2011nk}.
The LHC heavy ion experiments also confirmed the existence of this new state of matter
(see eg. Refs.~\cite{Aamodt:2010jd,Aamodt:2010pa,CMS:2012aa,Chatrchyan:2012ta}), meaning that at
two orders of magnitude higher collision energies, the created matter behaves very similarly, a result intriguing on its own.

The interest in relativistic hydrodynamics grew in past years mainly due to the above described
discovery of the nearly perfect fluidity of the experimentally created sQGP, and the success
of hydro models in describing the experimental data.
One of the most interesting open questions is perhaps whether collectivity also exists in
smaller systems, such as proton-nucleus or even proton-proton collisions.
Hydrodynamics can be successfully applied in describing systems of vastly different sizes 
(such as the expansion of the Universe and high energy particle and nuclear collisions),
as the fundamental equations contain no internal scale.
Hydrodynamical models utilize locally thermal distributions and apply the local conservation
of energy, momentum, and sometimes a conserved charge or entropy. Models based 
on hydrodynamics aim to describe the space-time picture of heavy-ion collisions
and infer the relation between experimental observables and the initial conditions.
The equations of hydrodynamics are highly nonlinear, so they are frequently solved
via simulations, where suitable initial conditions have to be assumed, and the
equations of hydrodynamics are then solved numerically, see details for example in the recent
reviews of Refs.~\cite{deSouza:2015ena,Huovinen:2013wma}. Besides these efforts, there is also an interest
in models where exact, explicit and parametric solutions of the hydrodynamical equations are used,
and where the initial state may be inferred directly from matching the parameters of the solution to the data.
Several famous hydrodynamical solutions were developed to describe high
energy collisions~\cite{Landau:1953gs,Hwa:1974gn,Bjorken:1982qr}, but also
advanced relativistic solutions were found in the last
decade~\cite{Csorgo:2003ry,Csorgo:2006ax,Csorgo:2007ea,Nagy:2007xn,
Csanad:2012hr,Borshch:2007uf,Pratt:2008jj,Csanad:2014dpa},
when a revival of this sub-field was seen.

While it is customary to describe the medium created in heavy ion collisions 
with hydrodynamic models,  proton-proton collisions are frequently
considered to form a system that might be too small for thermalization, or that might be
not hot or dense enough to create a supercritical (non-hadronic) medium.
Energy densities in $\sqrt{s}=200$ GeV p+p collisions are likely below this limit. 
It is however an interesting question, how high energy densities can be
reached in p+p collisions at the LHC with  $\sqrt{s}=7$ and higher collision energies, as
at sufficiently high energy densities, the degrees of freedom may allow for a collective
description of the system.

We recall here that the application of hydrodynamical expansion to data analysis in high
energy p+p collisions is not an unprecedented or new idea, as Landau worked out
hydrodynamics for p+p collisions~\cite{Belenkij:1956cd}, and Bjorken also notes this
possibility in his paper~\cite{Bjorken:1982qr}
describing his energy density estimate.
It is also noteworthy that Hama and Padula assumed~\cite{Hama:1987xv} the formation of an
ideal fluid of massless quarks and gluons in
p+p collisions at CERN ISR energies of $\sqrt{s}$ = 53- 126 GeV.
Alexopoulos et al. used Bjorken's estimate to determine the initial energy
density of $\sim1.1\pm0.2$ GeV/fm$^3$ at the Tevatron in $\sqrt{s}$ = 1.8 TeV
p+$\overline{{\rm p}}$ collisions in the E735 experiment~\cite{Alexopoulos:2002eh},
while L\'evai and M\"uller argued~\cite{Levai:1991be}, that the transverse
momentum spectra of pions and baryons indicate the creation of a
fluid-like quark-gluon plasma in the same experiment at the same Tevatron
energies. However, these earlier works considered the quark-gluon
plasma as an ideal gas of massless quarks and gluons, while the RHIC
experiments pointed to a nearly perfect fluid of quarks where the speed of
sound ($c_s$) is measured to be $c_s \approx 0.35 \pm 0.05$~\cite{Adare:2006ti} that is significantly
different from that of a massless ideal gas of quarks and gluons,
characterized by a $c_s = 1/\sqrt{3} \approx 0.57$.
More recently, Wong used Landau hydrodynamics to predict multiplicities and rapidity densities in
proton-proton collisions~\cite{Wong:2008ex}.
Furthermore, Shuryak and Zahed proposed~\cite{Shuryak:2013ke} the application of hydrodynamics for high
multiplicity p+p and p+A collisions at CERN LHC. The ridge effect~\cite{CMS:2012qk,Padula:2011yk},
i.e. long range azimuthal correlations were observed in high multiplicity p+p and p+A
as well as in heavy ion reactions, which also points towards similarities in collectivity,
and perhaps the applicability of hydrodynamics can be extended to these systems.

In this paper
we apply the hydrodynamical solution of Refs.~\cite{Csorgo:2006ax,Csorgo:2007ea}
to describe the pseudorapidity distribution in p+p collisions at $\sqrt{s}=7$ and 8 TeV
and use the results of these hydrodynamical fits to estimate the initial energy
density in these reactions.

\section{Rapidity distributions from hydrodynamics}
Equations of  hydrodynamics can be described by the continuity of conserved charges and that of the
energy-momentum-tensor:

\begin{align}
\partial_{\nu} (nu^{\nu}) = 0,\qquad
\partial_{\nu}T^{\mu\nu} = 0,
\end{align}
with $n$ being a conserved charge, and $T$ is the energy-momentum tensor. In case of a perfect fluid
the latter can be expressed as

\begin{align}
T^{\mu\nu}=(\epsilon + p) u^{\mu}u^{\nu}-pg^{\mu\nu}.
\end{align}
where $u^\mu$ is the velocity field, $\epsilon$ is the energy density and $p$ the pressure, and this tensor is equal
to diag($\epsilon,-p,-p,-p$) in the locally comoving frame, where $u^\mu=(1,0,0,0)$. An Equation of State (EoS)
is needed to close the above set of equations, and for that, $\epsilon = \kappa p$ is frequently utilized with a
constant $\kappa$ value. Furthermore, the temperature $T$ may be defined as $p=nT$, or if there are no 
conserved charges, the relation $\epsilon+p=\sigma T$ (with $\sigma$ being the entropy density) may be utilized.
Note that if $\kappa$ is a constant, independently of the temperature, then it is simply
connected to the speed of sound ($c_s$) as $\kappa=1/c_s^2$.
An analytic hydrodynamical solution is a functional
form of $\epsilon$, $p$, $T$, $u^\mu$ and $n$, which solves the above equations. 
The solution is explicit if these fields are explicitly expressed as a function of
space-time coordinates $x^\mu=(t,\mathbf{r})=(t,r_x,r_y,r_z)$.

Let us utilize the Rindler-coordinates, where $\tau=\sqrt{t^2-{\mathbf r}^2}$ is a coordinate proper-time,
$\eta_S=0.5\sqrt{(t+|\mathbf{r}|)/(t-|\mathbf{r}|)}$ the space-time rapidity. With these, we can
discuss the solution detailed in Refs.~\cite{Csorgo:2006ax,Csorgo:2007ea,Nagy:2007xn,Csorgo:2008pe} as:

\begin{gather}
u^\mu = ({\rm ch}\lambda\eta_S,{\rm sh}\lambda\eta_S),\qquad
n =  n_f\frac{\tau_f^{\lambda}}{\tau^{\lambda}},\qquad
T =  T_f\left(\frac{\tau_f}{\tau}\right)^{\frac{\lambda}{\kappa}}\label{e:sol3},
\end{gather}
where subscript $f$ denotes
quantities at the freeze-out, while $\lambda$ controls the acceleration.
If $\lambda=1$, there is no acceleration
and (if the expansion is one-dimensional) we get back the accelerationless Hwa-Bjorken solution
of Ref.~\cite{Hwa:1974gn,Bjorken:1982qr}.
For $\lambda>1$, we obtain several classes of accelerating solutions, described 
 in Refs.~\cite{Csorgo:2006ax,Csorgo:2007ea,Nagy:2007xn,Csorgo:2008pe}. For example,
1+1 dimensional hydrodynamical solutions are obtained for any real value of $\lambda$, for the
special EoS of $\kappa=1$. Also $\kappa=d$ (with $d$ being the number of dimensions) solutions
can be obtained with $\lambda=2$, other possibilities are summarized in the above references. We 
will use the 1+1 dimensional solution, as this can be applied well to describe the longitudinal dynamics
of the system, and to estimate the initial energy density based on final state observables, similarly
to the original paper of Bjorken~\cite{Bjorken:1982qr}. Even though this solution is valid with the
EoS of $\kappa=1$, we will later see that we can make a simple additional correction to make up
for this shortcoming of the given solution. Also note that in Ref~\cite{Nagy:2007xn}
an exact 1+1 dimensional solution was also given for fluctuating initial conditions, so in this case the
$\kappa=1$ case is solved for arbitrary initial conditions. This also allowed to confirm the stability of the
solutions.

To apply the above solutions of hydrodynamics, one has to calculate hadron momentum distributions.
In order to do so, one has to utilize a freeze-out condition. Let us define $T_f$ as the freeze-out temperature
at $\eta_S=0$, and the freeze-out hypersurface shall be pseudo-orthogonal to the velocity field, i.e.
$u^\mu(x) \parallel {\rm d}\Sigma_\mu(x)$ where ${\rm d}\Sigma_\mu(x)$ is the vector-measure of the
freeze-out hypersurface. In the case of the above discussed solution, the equation of the hypersurface
will be $(\tau_f/\tau)^{1-\lambda}=\cosh((\lambda-1)\eta_S)$. Using this, one may calculate rapidity distribution $dN/dy$
(with $N$ being then the total number of hadrons (or that of charged hadrons), and
$y=0.5 \ln \left((E+p_z)/(E-p_z)\right)$ the rapidity). This was performed
in Refs.~\cite{Csorgo:2006ax,Csorgo:2007ea,Nagy:2007xn,Csorgo:2008pe}, and the approximate
analytic result is:

\begin{align}\label{e:dndy-approx}
\frac{dN}{dy}\approx N_0 \cosh^{\frac{\alpha}{2}-1}\brackets{\frac{y}{\alpha}}
e^{-\frac{m}{T_f}\cosh^\alpha\brackets{\frac{y}{\alpha}}} ,
\end{align}
with $\alpha=\frac{2\lambda-1}{\lambda-1}$ containing the acceleration parameter $\lambda$,
$m$ is the average mass of the hadrons to be described (typically
this is very close to the pion mass) and $N_0$ is a normalization parameter, to be determined by fit to the data.
Note that here we defined the four-momentum components as $p^\mu=(E,p_x,p_y,p_z)$.
The rapidity distribution is approximately Gaussian with the width of $ \alpha/(m/T_f + 1/2 + 1/\alpha)$,
where $m$ is the mass of the particles for which we calculate the rapidity density. 
At $\lambda=1$, the width becomes infinity and the distribution becomes flat, as in the Bjorken limit
(corresponding to the Hwa-Bjorken solution), where the solution becomes boost invariant.
Note that fluctuating initial conditions in our case would result in fluctuations of the measured
rapidity densities. However, if the fitted data are smooth, the final state fluctuations are averaged out,
so the initial condition can be taken as a smooth distribution as well.

It is important to observe that in order to describe the majority of experimental data, besides rapidity
distributions, pseudorapidity distributions (with pseudorapidity defined as $\eta=0.5 \ln \left((p+p_z)/(p-p_z)\right)$) have
to be calculated as well~\cite{Csorgo:2006ax,Csorgo:2007ea,Nagy:2007xn,Csorgo:2008pe}.
This can be done by using an average transverse momentum ($p_t$) value
and making a transformation from pseudorapidity $\eta$ to rapidity $y$.
The double differential particle number distributions are connected as 

\begin{align}
\frac{E}{p}\frac{1}{p_t}\frac{dN}{dp_td\eta}=\frac{1}{p_t}\frac{dN}{dp_t dy}
\end{align}
thus the we obtain

\begin{align}
\frac{dN}{d\eta}=\frac{\bar{p}_T\cosh\eta}{\sqrt{m^2+\bar{p}_T^2\cosh^2\eta}}\frac{dN}{dy}
\end{align}
with $\bar{p}_T$ being the (rapidity dependent) average transverse
momentum of these particles. The value of latter can be estimated from the effective temperature ($T_{\rm eff}$)
of hadron spectra, e.g. using the Buda-Lund modell~\cite{Csorgo:1995bi,Csanad:2003qa}, that indicates a behavior
observed already in $\sqrt{s} = 22$ GeV h+p collisions at the EHS/NA22 Collaboration~\cite{Agababyan:1997wd},
and is a generic property of 3 dimensional, finite, exact hydrodynamical solutions with directional
Hubble flows~\cite{Csorgo:2001xm,Csorgo:2003ry,Csanad:2009wc}. These findings can be summarized as:

\begin{align}
\bar{p}_T&=\frac{T_{\rm eff}}{1+\sigma^2y^2}\textnormal{ with }\\
\sigma^2&=\frac{T_fT_{\rm eff}}{m^2\left(\Delta y^2+T_f/m\right)}\textnormal{ and }\\
T_{\rm eff}& = T_f + \frac{m\langle u_t \rangle^2}{1+m/T_f}
\end{align}
and here $T_f$, $\langle u_t \rangle$ and $\Delta y$ are model parameters describing the central
temperature, the average transverse flow and the rapidity distribution width, all at the freeze-out.
These can be extracted from model comparisons, and here we made the simplification of using just the
first formula of the above equations and using $T_{\rm eff}$ and $\sigma$ as the only model parameters,
where the first was fixed to a value of 170 MeV, while the second was determined by the fits.

\section{Energy density estimation}
In this section we recapitulate how this model can be used for
improving the famous energy density estimation made by Bjorken~\cite{Bjorken:1982qr}. 
Let us look at the thin transverse slab at mid-rapidity, at the
coordinate proper-time of the thermalization $\tau=\tau_0$, which is in fact the initial time of a possible hydro type of evolution.
This thin slab is illustrated by Fig. 2 of Ref.~\cite{Bjorken:1982qr}.
The size of this slab is estimated by the radius  $R$ of the colliding hadrons or nuclei, them the initial
volume is $dV=(R^2\pi)\tau_0 d\eta_0$, with $\tau_0 d\eta_{S0}$ being the longitudinal size, while
$d\eta_{S0}$ is the space-time rapidity width at $\tau_0$, as detailed in
Refs.~\cite{Csorgo:2006ax,Csorgo:2007ea,Nagy:2007xn,Csorgo:2008pe}.
The energy contained in this volume is $dE = \langle E\rangle dN$, where $dN$ is the number of
particles and $\langle E \rangle$ is their average energy near $y=0$. 
In the special case of the accelerationless, boost-invariant Hwa-Bjorken flow, $\eta_{S0}$ equals
to the the freeze-out width of the same slab, $\eta_{Sf}$, as illustrated by
Fig. 4.2 of Ref.~\cite{Nagy:2012fma}. Furthermore, due to boost-invariance, $\eta_{Sf}=y$
symbolically.
With this, Bjorken concludes
on the initial energy density for a boost invariant solution:

\begin{align}\label{e:Bjorken}
    \epsilon_{\rm Bj} = \frac{\langle E\rangle dN}{(R^2 \pi)\tau_0d\eta_0} =
\frac{\langle E\rangle}{(R^2 \pi)\tau_0}\left.\frac{dN}{dy}\;\right|_{y=0} .
\end{align}

However, in case of an accelerating solution, the latter two assumptions do not hold,
hence one arrives at a correction to take into
account the acceleration effects on the energy density estimation,
as discussed e.g. in Ref.~\cite{Nagy:2007xn}.
This correction can be calculated from the partial derivatives

\begin{align}
\frac{\partial y}{\partial\eta_{Sf}}\frac{\partial\eta_{Sf}}{\partial\eta_{S0}}=
\brackets{2\lambda-1}\brackets{\frac{\tau_f}{\tau_0}}^{\lambda-1}
\end{align}
Thus for the hydrodynamical solutions where the acceleration parameter is $\lambda>0$.
the initial energy density is given by a corrected  estimation $\epsilon_{\rm corr}$ as 

\begin{align}\label{e:ncscs}
\epsilon_{\rm corr}=\epsilon_{\rm Bj}\brackets{2\lambda-1}\brackets{\frac{\tau_f}{\tau_0}}^{\lambda-1}
\end{align}
Here $\epsilon_{\rm Bj}$ is the Bjorken estimation, which is recovered if $dN/dy$ is flat
(i.e. $\lambda=1$), but for $\lambda>1$, both correction factors are bigger than 1.
These correction factors take into account the work done by the pressure on the surface
of a finite and accelerating, hot fireball.  Hence
the initial energy densities are under-estimated by the Bjorken
formula, if the measured $dN/dy$ distributions are not constant, but have a finite width.
In Refs.~\cite{Csorgo:2006ax,Csorgo:2007ea,Nagy:2007xn,Csorgo:2008pe} we performed fits
to BRAHMS pseudo-rapidity distributions from Ref.~\cite{Bearden:2001qq}, and these fits
indicate that $\epsilon_{\rm corr}\approx 10$ GeV/fm$^3$ in $\sqrt{s_{\rm NN}}=200$ GeV
Au+Au collisions at RHIC.

The above corrections are exact results derived in details for a special EoS
of $\kappa =1$~\cite{Nagy:2007xn}. However, the relation of the pressure to the energy
density is obviously EoS dependent. As of today, no simple, exact, analytic solutions of the
equations of relativistic hydrodynamics are known that utilize other EoS values. However,
as proposed in Refs.~\cite{Csorgo:2008pe,Nagy:2012fma}
the effects of a non-ideal EoS can be estimated with a conjecture. Any result about the
EoS dependent initial energy density shall fulfill the following simple requirements:
\begin{enumerate}
\item It has to reproduce the EoS-independent Bjorken estimate for $\lambda\rightarrow 1$.
\item It has to reproduce the exact result of Eq.~(\ref{e:ncscs}) for any $\lambda$,
        in the $\kappa\rightarrow 1$ limit.
\item It has to follow known hydro behavior for $\epsilon(\tau)$ corresponding to exact solutions valid
        for any (temperature independent) EoS, see e.g. Refs.~\cite{Csorgo:2003rt,Csorgo:2001xm},
        where $\epsilon\propto(\tau_0/\tau)^{1/\kappa}$ behavior is found, assuming a fluid volume
        proportional to $\tau$.
\item It should approximately reproduce the results of numerical hydro calculations, most importantly
        the additional correction for $\kappa>1$ should increase the initial energy density.
\end{enumerate}
By the principle of Occam's razor, we then arrive at

\begin{align}
\epsilon_{\rm corr} = \epsilon_{\rm Bj}\brackets{2\lambda-1}\brackets{\frac{\tau_f}{\tau_0}}^{\lambda-1}
\brackets{\frac{\tau_f}{\tau_0} }^{(\lambda-1)\left(1-\frac{1}{\kappa}\right)} \label{e:conjeps} 
\end{align}
This conjecture satisfies the previously mentioned consistency requirement. It goes back
to the initial energy density of Bjorken in the exactly solvable $\lambda = 1$ special case (for any $c_s$),
and it also gives the correct energy density for $\lambda \neq 1$ for the exactly solvable $\kappa = 1$ special case.
Furthermore, this estimate was cross-checked against numerical solutions of relativistic hydrodynamics, that
reproduce rapidity distributions.  For example Ref.~\cite{Schenke:2010nt}, using various EoS versions,
obtains 55 GeV/fm$^3$ in 200 GeV Au+Au collisions for 0.4 fm$/c$ or approximately 35-40 GeV/fm$^3$
at 0.55 fm$/c$. Evaluation of these values for $\tau_0 = 1$ fm/$c$ (assuming 3D expansion and an EoS
value valid at high temperatures) we obtain $\approx$20 GeV/fm$^3$ initial energy density, which is
similar to the value of 15 GeV/fm$^3$ that is obtained from our conjectured formula in
Ref.~\cite{Csorgo:2008pe,Nagy:2012fma}.

From basic considerations~\cite{Bjorken:1982qr}, as well as from lattice QCD calculations~\cite{Borsanyi:2010cj},
it follows that the critical energy density, needed to form a non-hadronic medium is around 1 GeV/fm$^3$. From the
lattice QCD calculations one gets $\epsilon_{\rm crit}=(6-8)\times T_{\rm crit}^4$ (in $\hbar c=1$ units), and
even with a conservative estimate of $T_{\rm crit} = 170$ MeV, one gets $\epsilon_{\rm crit}<1$ GeV/fm$^3$.
Thus initial energy densities above this value indicate the formation of a non-hadronic medium, which, in
case of RHIC Au+Au collisions at 200 GeV/nucleon collision energy, is confirmed by the observations 
of the RHIC White Papers~\cite{Adcox:2004mh,Adams:2005dq,Back:2004je,Arsene:2004fa}.
It is however an interesting question, as already addressed in the introduction, whether a non-hadronic
medium may also be formed in small systems, such as p+p collisions. The RHIC energies are most
probably not high enough for that, so let us investigate LHC p+p collisions in the next section.

\section{Initial energy densities in 7 and 8 TeV LHC p+p collisions}\label{s:totemfit}
Now us utilize the Bjorken estimate first to calculate the initial energy density in 7 TeV p+p collisions.
For that, we need to estimate the quantities indicated in Eq.~(\ref{e:Bjorken}).
The average transverse momentum in $\sqrt{s}=7$ TeV p+p collisions is
$\langle p_t\rangle = 0.545 \pm 0.005_\textnormal{stat}\pm 0.015_\textnormal{syst}$ GeV$/c$~\cite{Khachatryan:2010us},
which corresponds to  $\langle E\rangle=0.562$ GeV$/c^2$ at midrapidity
(assuming most of these particles are pions). 

It is a non-trivial question, how to estimate the initial transverse area $R^2\pi$ in p+p collisions,
because the geometrical area relates to the total, elastic, inelastic  and differential cross-sections
in an involved and non-trivial manner. In case of the collisions of large heavy ions, the initial overlap 
area is evaluated based on nuclear density profiles, for example using the relation where $R^3$ is 
proportional to the atomic number of the given nucleus. Basically these relations that determine
the nuclear geometry were obtained from the analysis of the differential cross-sections
of elastic electron-ion~\cite{Hofstadter:1956qs} and elastic proton-nucleus data~\cite{Glauber:1970jm}. 
Similarly, to get a reliable estimate of the initial transverse area in p+p collisions,
we should rely on the analysis of elastic p+p scattering data.

Our analysis is based on Eqs. (117-119) and (124-126) of Ref.~\cite{Block:2006hy},
that show that both for a grey disc and for a Gaussian scattering density profile,
$\sigma_{\rm el} = \pi R^2 A^2$ and $\sigma_{\rm tot} = 2 \pi R^2 A$, where $A$ measures
the ``greyness" of the proton. Thus the geometrical area $R^2 \pi$ actually can be estimated as

\begin{align}
\pi R^2 =\frac{\sigma_{\rm tot}}{2 A} = \frac{\sigma_{\rm tot}^2}{4  \sigma_{\rm el}}.
\end{align}
We have cross-checked these estimates by evaluating the geometrical area from $B$, the slope
of differential elastic scattering cross-section at zero momentum transfer, as we may also use the
relation $R^2\pi  = 4\pi B$~\cite{Block:2006hy}. We found that within errors both methods yield the
same estimate for the initial geometry of p+p collisions. Based on the results of
Refs.~\cite{Antchev:2011zz,Antchev:2013iaa}, we have conservatively estimated $R=1.76\pm0.02$ fm.


Furthermore, the formation time, $\tau_0$, may conservatively assumed to be 1 fm$/c$, as
was done in Bjorken's paper as well. The only remaining parameter is 
the multiplicity or (pseudo)rapidity density at midrapidity. As measured by the LHC experiments, the charged particle
pseudorapidity density at midrapidity is found to be
$6.01\pm0.01\textnormal{(stat)}^{+0.20}_{-0.12}\textnormal{(syst)}$ at ALICE~\cite{Khachatryan:2010us},
while $5.78\pm0.01_\textnormal{stat}\pm0.23_\textnormal{syst}$ at CMS~\cite{Aamodt:2010pp}, but in
some multiplicity classes it may reach values of 25-30 (see Table I. of Ref.~\cite{Aamodt:2011kd}). We
will take the average of the first two values. The total multiplicity is then 3/2$\times$ the charged
particle multiplicity. Substituting all of the above mentioned values to Eq.~(\ref{e:Bjorken}), one gets:

\begin{align}
  \epsilon_{Bj}(7\;{\rm TeV}) =
  0.507\;\rm{GeV/fm}^3,
\end{align}
which is below the critical value.

\begin{figure}
\begin{center}
\includegraphics[width=0.7\linewidth,clip]{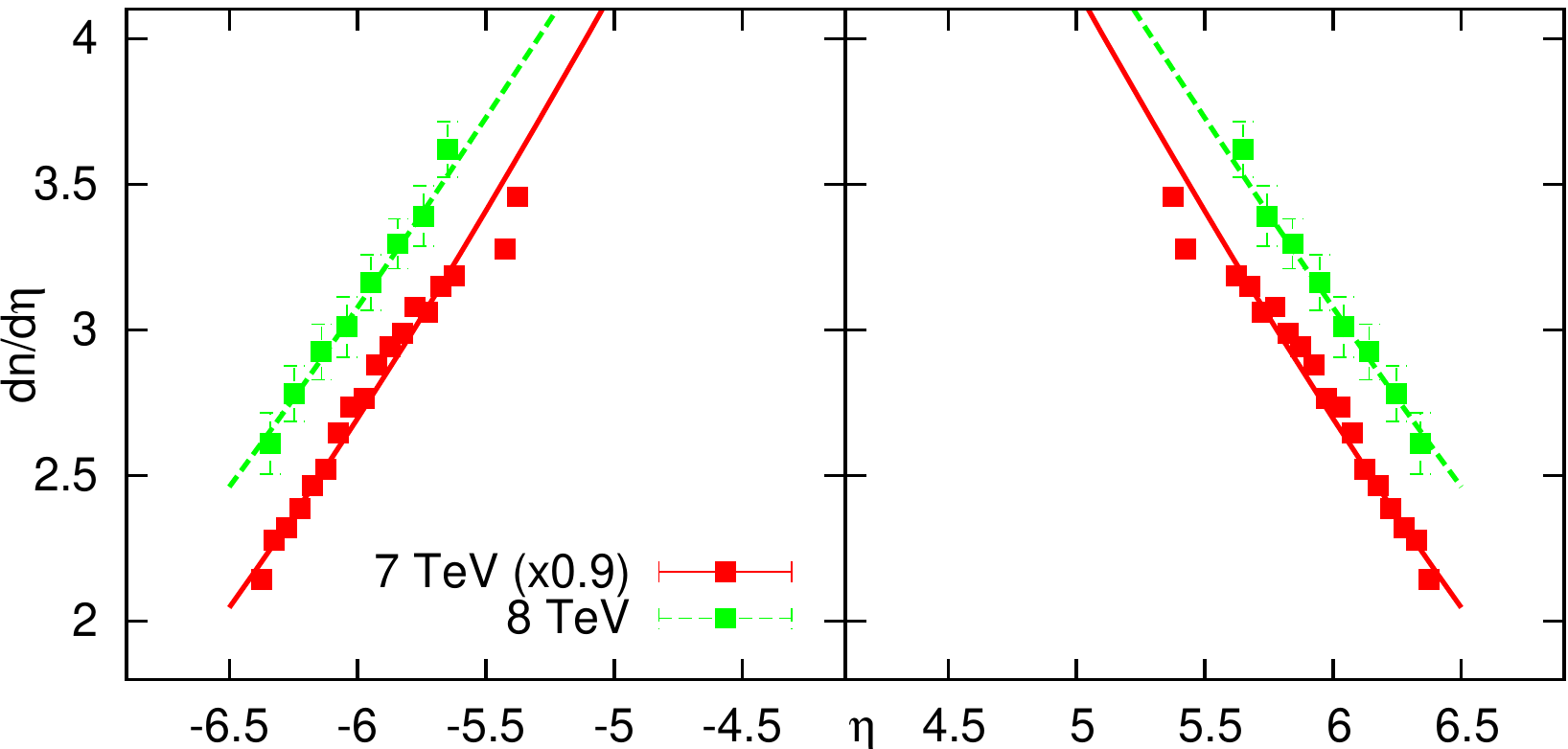}
\end{center}
\caption{\label{f:totemfit} Charged particle
$\frac{dN}{d\eta}$ distributions from TOTEM
fitted with the result of the relativistic hydro solution described
in this paper.}
\end{figure}

Let us now make an advanced estimate of the initial energy density. Such an estimate may 
be based on TOTEM pseudorapidity density data, as these reach
out to large enough $\eta$ values so that the acceleration parameter can be determined.
Fits to TOTEM data were performed via Eq.~(\ref{e:dndy-approx}), as shown in 
Fig.~\ref{f:totemfit}. The fit resulted in the acceleration parameter
$\lambda = 1.073\pm0.001_\textnormal{stat}\pm0.003_\textnormal{syst}$,
where the systematic error is based on the point-to-point systematic error of the data points.
The fit also determined the normalization parameter $N_0$ to be 7.45 (with a systematic
uncertainty of approximately 3\%, contributing to the initial energy estimate),
and the $dN/dy\rightarrow dN/d\eta$ conversion parameter $\sigma$ to be 0.8, with an
uncertainty contributing to the initial energy density mainly through the value of $\lambda$.
We can now use the $\lambda$ value to make a realistic estimation of the initial energy density. 

\begin{figure}
\begin{center}
\includegraphics[width=0.7\linewidth,clip]{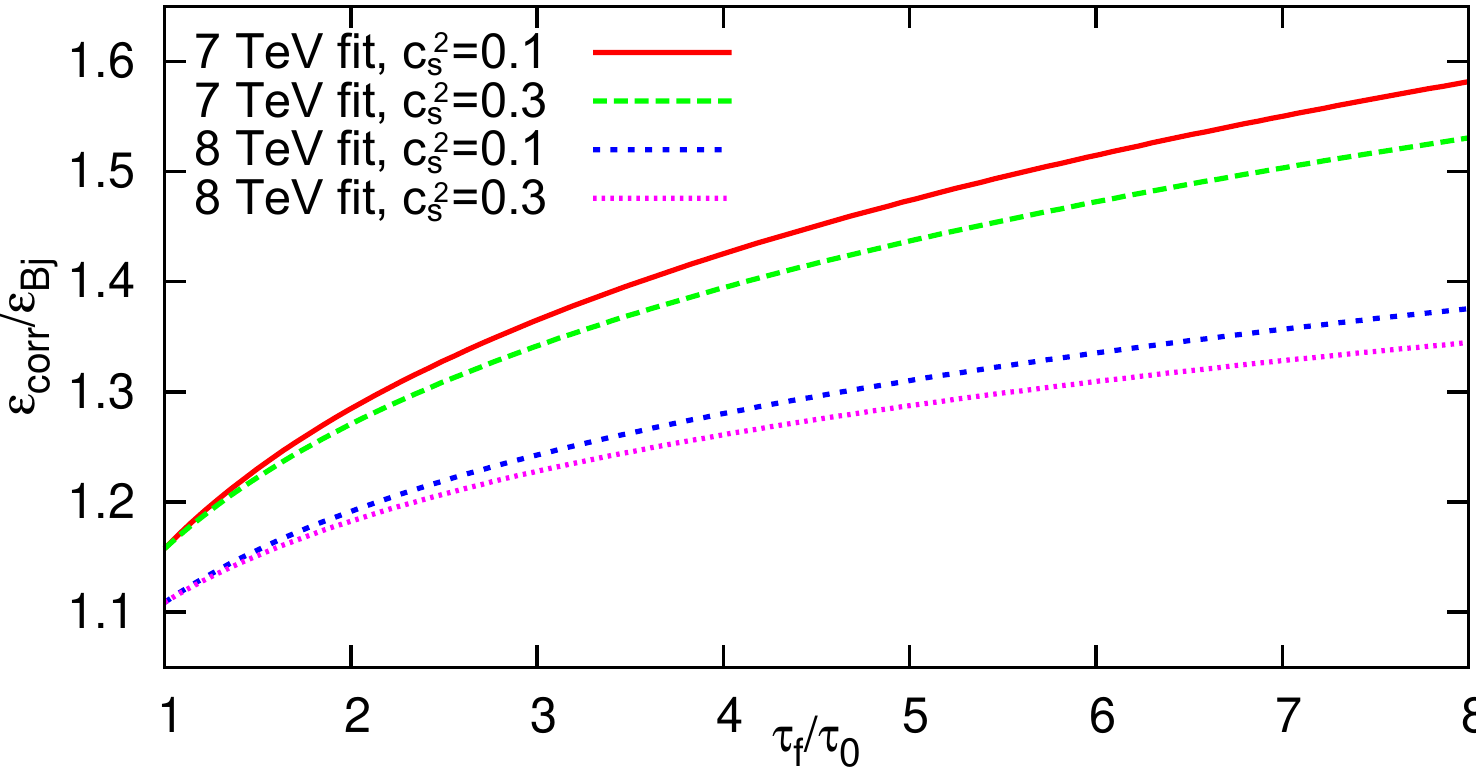}
\end{center}
\caption{\label{f:cscorr} The correction factor as a function of
freeze-out time versus thermalization time ($\tau_f/\tau_0$). At a
reasonable value of ~2, the correction factor is around 25\%.}
\end{figure}

Assuming $c_s^2=1/\kappa=0.1$ (this is a quite realistic value, at least no harder EoS is expected at LHC,
as similar EoS was found at RHIC as well~\cite{Lacey:2006bc,Borsanyi:2010cj,Csanad:2011jq}), one
only needs a $\tau_f$ value. As shown in Eq.~(\ref{e:sol3}), temperature is proportional to
$\tau^{-\lambda/\kappa}$.
From this, $\tau_0 = \tau_f (T_f/T_0)^{\kappa/\lambda}$. Thus if the freeze-out temperature is $T_f=140$ MeV,
then an initial temperature of $T_0 = 170$ MeV (needed in order to form a strongly interacting quark
gluon plasma) corresponds to $\tau_f$ being 5-6 times $\tau_0$, for $c_s^2=0.1$
and $\lambda=1.1$. Even if $c_s^2$ and $\lambda$ are higher, $\tau_f/\tau_0$ seems to
be a rather conservative value. With this, one gets a correction factor of 1.262, thus
the corrected initial energy density estimate is

\begin{align}\label{e:corrfact}
\epsilon_{\rm corr}(7\;{\rm TeV}) = 
0.640\;\rm{GeV/fm}^3,
\end{align}
which is still below the critical value. The $c_s^2$ and $\tau_f/\tau_i$ dependence of the
correction factor is shown in Fig.~\ref{f:cscorr}.

\begin{figure}
\begin{center}
\includegraphics[width=0.95\linewidth]{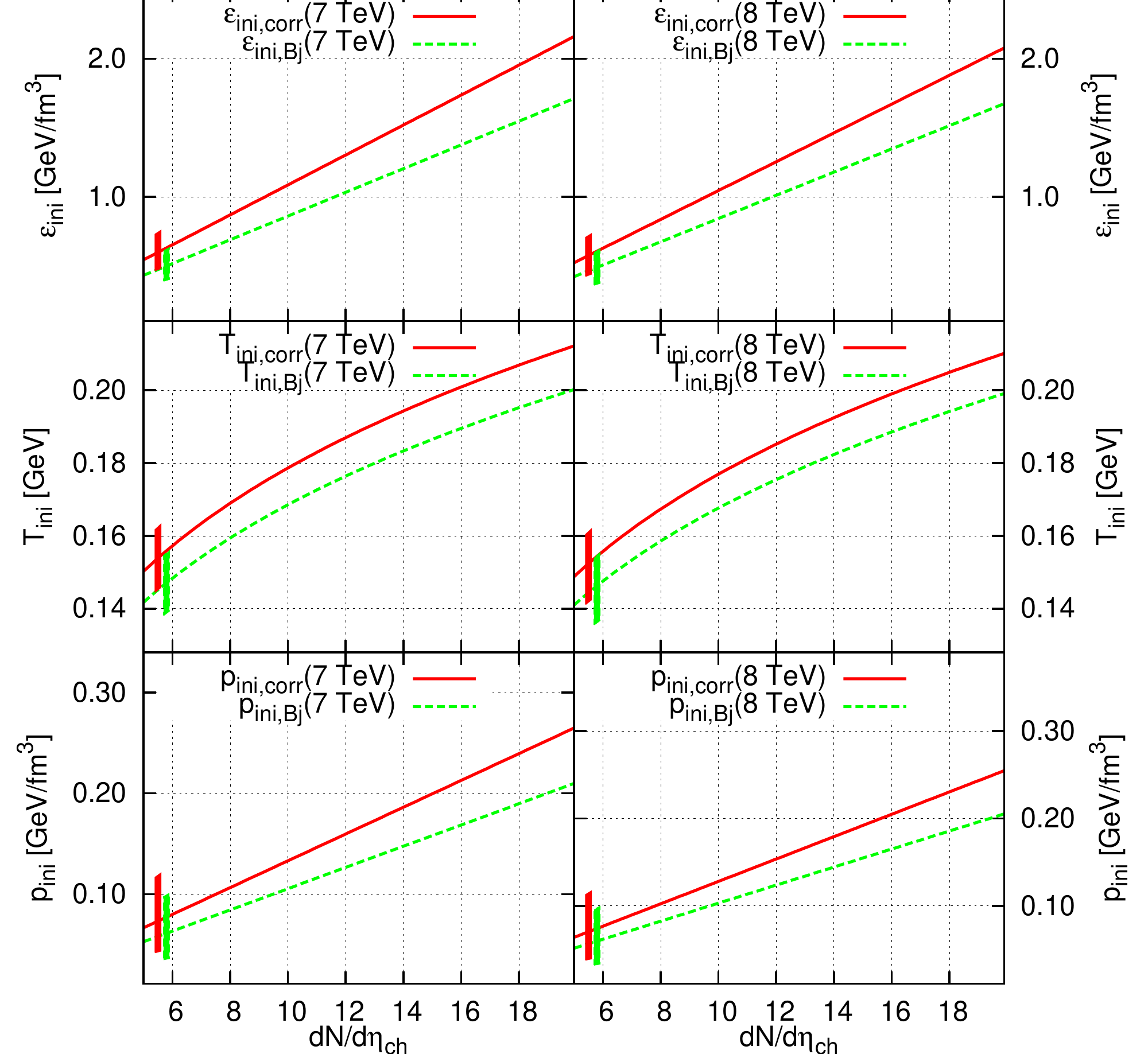}
\end{center}
\caption{\label{f:multdep}Initial energy density (based on Table~\ref{t:errors}), temperature
and pressure (based on the $\epsilon\propto T^4$ relationship) at 7 TeV (left) and 8 TeV (right),
is indicated as a function of central charged particle multiplicity density.
The Bjorken-estimate (dashed curve) is above the critical energy density of 1 GeV/fm$^3$ if the
multiplicity is larger than 12. Corrected initial energy density (solid curve) is above
the critical value if multiplicity is larger than 9. Boxes (parallelograms) show systematic
uncertainty (estimated from the 7 TeV case).}
\end{figure}

Note that the average p+p multiplicity was used here, so
this value represents an average energy density in p+p collisions -- below
the critical value of 1 GeV/fm$^3$ (c.f. Ref~\cite{Campanini:2011bj}
where a possible cross-over starting at $dN/d\eta|_{\eta=0}=6$ was conjectured). Based on Table I. of
Ref.~\cite{Aamodt:2011kd}, much larger multiplicities have been reached however. The energy
density results for these multiplicities are shown on Fig. \ref{f:multdep}. It is clear from this
plot, that even for the original Bjorken estimate, supercritical energy densities may have
been reached in high multiplicity events, roughly above a charged particle multiplicity of 12.
The corrected estimate gives supercritical values for charged particle multiplicities of 9.
We also calculated the initial temperature based
on the  $\epsilon\propto T^4$ relationship, assuming that 175 MeV corresponds to 1 GeV/fm$^3$
approximately. This is also shown in the left panel of Fig.~\ref{f:multdep}, as well as the
reachable pressure values. 
It is known that temperature of 300-600 MeV may have been reached in 200 GeV central Au+Au collisions of
RHIC~\cite{Adare:2008ab}. Initial temperature values in 7 TeV p+p seem to be lower than that,
300 MeV can be reached in events with a multiplicity of $>50$. However, a temperature of
200 MeV may already be reached in events with a multiplicity of 16.

Now let us estimate what happens at $\sqrt{s}=8$ TeV. As for the Bjorken-estimate, we need the change
in charged particle multiplicity, average transverse energy and transverse size. CMS indicates
$dN/d\eta|_{\eta=0}=6.20\pm0.46$ for a non-single diffractive enhanced data sample in
Ref.~\cite{Chatrchyan:2014qka}, while $dN/d\eta|_{\eta=0}=6.13\pm0.1$ is given in
Ref.~\cite{Adam:2015gka}. The average of the two values is in good
agreement with the approximate $s$ dependence of $dN/d\eta|_{\eta=0}$ of $0.715\cdot \sqrt{s}^{0.23}$
as estimated in Ref.~\cite{CMS:2013yca}.
Average transverse momentum $s$ dependence is estimated as
$\langle p_t \rangle =0.413 - 0.0171 \ln s + 0.00143 \ln^2 s$
in Ref.~\cite{Khachatryan:2010us}, which means a 1.53\% increase in $\langle E \rangle$.
Transverse size can be estimated based on the $\sigma_{\rm tot}$ measurement of
Refs.~\cite{Antchev:2013paa}, $\sigma_{\rm tot}=101.7\pm2.9{\rm(syst)}$ mb, which means
a 3.8\% increase in area compared to 7 TeV, and $R=1.799\pm0.025$ fm
Based on Eq.~(\ref{e:Bjorken}), this altogether means a
2.4\% increase in $\epsilon_{\rm Bj}$, i.e. 

\begin{align}
\epsilon_{Bj}(8\; {\rm TeV}) =
 0.519\;\rm{GeV/fm}^3,
\end{align}
which is again below the critical value.
We also fitted $dN/d\eta$ data from TOTEM~\cite{Chatrchyan:2014qka}
as shown in Fig.~\ref{f:totemfit}. We obtained $\lambda = 1.067\pm 0.001$ in this case.
This corresponds to a correction factor of 1.240, similarly to Eq.~(\ref{e:corrfact}).
Finally, we arrive at

\begin{align}\label{e:8tev}
\epsilon_{\rm corr}(8\;{\rm TeV}) = 
0.644\;\rm{GeV/fm}^3.
\end{align}
This value is based on the average multiplicity in $\sqrt{s}=8$ TeV collisions.
However, at a fixed multiplicity, there is almost no difference between the two collision
energies: average transverse energy increases by 1.5\%, but the transverse size also
increases by 2.4\%. This means a roughly 1\% decrease, which is much smaller than the
systematic uncertainties in this estimate -- to be discussed in the next section. We plot the
multiplicity dependence of $\epsilon_{\rm ini}$, $T_{\rm ini}$ and $p_{\rm ini}$ for the 8 TeV
case in the right panel of Fig.~\ref{f:multdep}. We may again observe, that supercritical
values are reached for multiplicities higher than 10 in  case of the corrected estimate; but
even Bjorken's estimate yields supercriticality if $dN/d\eta|_{\eta=0}>13$.

\section{Uncertainty of the estimate}
Different sources of uncertainties are detailed in Table~\ref{t:errors}. The most important one comes
from $dN/d\eta$ at midrapidity. From Fig.~\ref{f:multdep} it is clear that for the Bjorken-estimate,
energy density is above the critical value of 1 GeV/fm$^3$ if the multiplicity is larger than 12,
while in case of the corrected initial energy density, $dN/d\eta|_{\eta=0}>9$ is needed.
Taking all sources of uncertainties into account, the final result for the energy density corresponding to mean
multiplicity density at 7 TeV is

\begin{align}
\epsilon_{\rm corr}(7\;{\rm TeV}) = 0.64 \pm 0.01\rm{(stat)}^{+0.14}_{-0.10}\rm{(syst)}\;\rm{GeV}/\rm{fm}^3
\end{align}
and the main systematic error comes from the estimation of the ratio $\tau_f/\tau_0$. In the 8 TeV case,
the estimate yields a somewhat larger number (0.644 versus 0.640 GeV/fm$^3$), but the uncertainties
are somewhat higher due to extrapolations to 8 TeV.

An important source of systematic uncertainty is the use of the given hydrodynamic solution.
This uncertainty may be estimated by using other hydrodynamic models that contain acceleration:
the Landau model~\cite{Landau:1953gs}, the Bialas-Peschanski model~\cite{Bialas:2007iu},
or numeric models of hydrodynamics, however, in the current paper we focus on the analytic
results that can improve on Bjorken's famous initial energy density
estimate. Let us also note, that we describe the measured pseudorapidity distributions by
suitably choosing i.e. fitting the initial conditions. With this, we predict only a moderate
correction to the Bjorken estimate of the initial energy density. However, in case of
numerical solutions, one has to assume a distribution of initial conditions, and one cannot
directly determine the initial conditions from the data.
A more detailed, or a numerical hydrodynamical investigation is
outside the scope of the present manuscript, however we can cross-check these
results by fitting the simultaneously taken CMS and TOTEM pseudorapidity density data
with the same model, to investigate the stability of our initial energy density
estimate from the details of pseudorapidity density at midrapidity.

\begin{table}
  \caption{Sources of statistical and systematic errors for the 7 TeV estimate.\label{t:errors}}	
  \centering
\begin{tabular}{|c|c|c|c|}
\hline
  parameter                        & value & stat. & syst. eff. on $\epsilon$ \\
\hline
  $\lambda$                        & 1.073 & 0.1\% & 0.4\% (from data) \\
  $c_s^2$                          & 0.1   &  -    & -2\%+0.2\% (if $0.05<c_s^2<0.5$)\\
  $\tau_f/\tau_0$ & 2              &  -    & -4\%+10\% (for $\tau_f/\tau_0$ in 1.5--4) \\
  $\tau_0$  [fm$/c$]               & 1     &  -    & underestimates $\epsilon$ \\
  $R$   [fm]                       & 1.766 &  -    & 1.3\% (from $\sigma_{\rm tot}$)\\
  $\langle E \rangle$ [GeV$/c^2$]  & 0.562 & 0.5\% & 3\%\\
  $dN/d\eta|_{\eta=0}$               & 5.895 & 0.2\% & 3\% (equivalently from $N_0$)\\
\hline
  \end{tabular}
\end{table}

\section{Improved initial energy density using combined TOTEM+CMS $dN/d\eta$ data}

\begin{figure}
\includegraphics[width=0.5\linewidth]{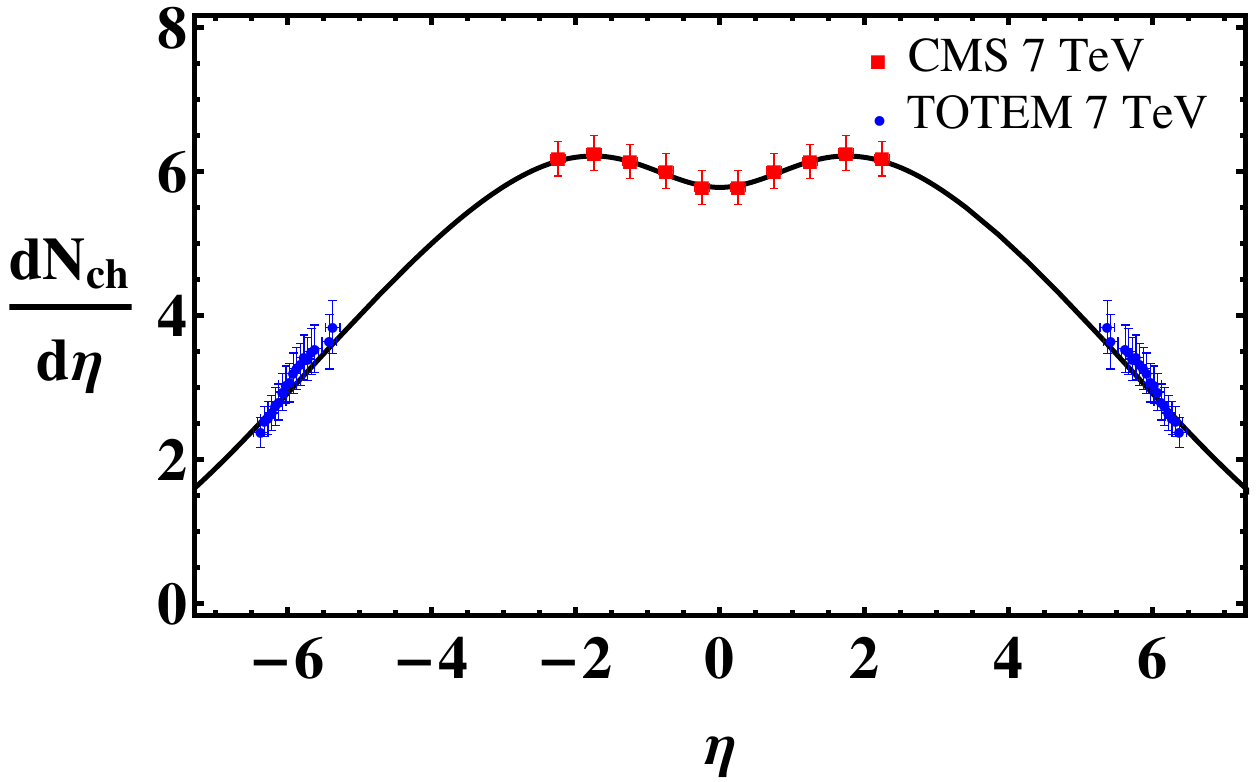}
\includegraphics[width=0.5\linewidth]{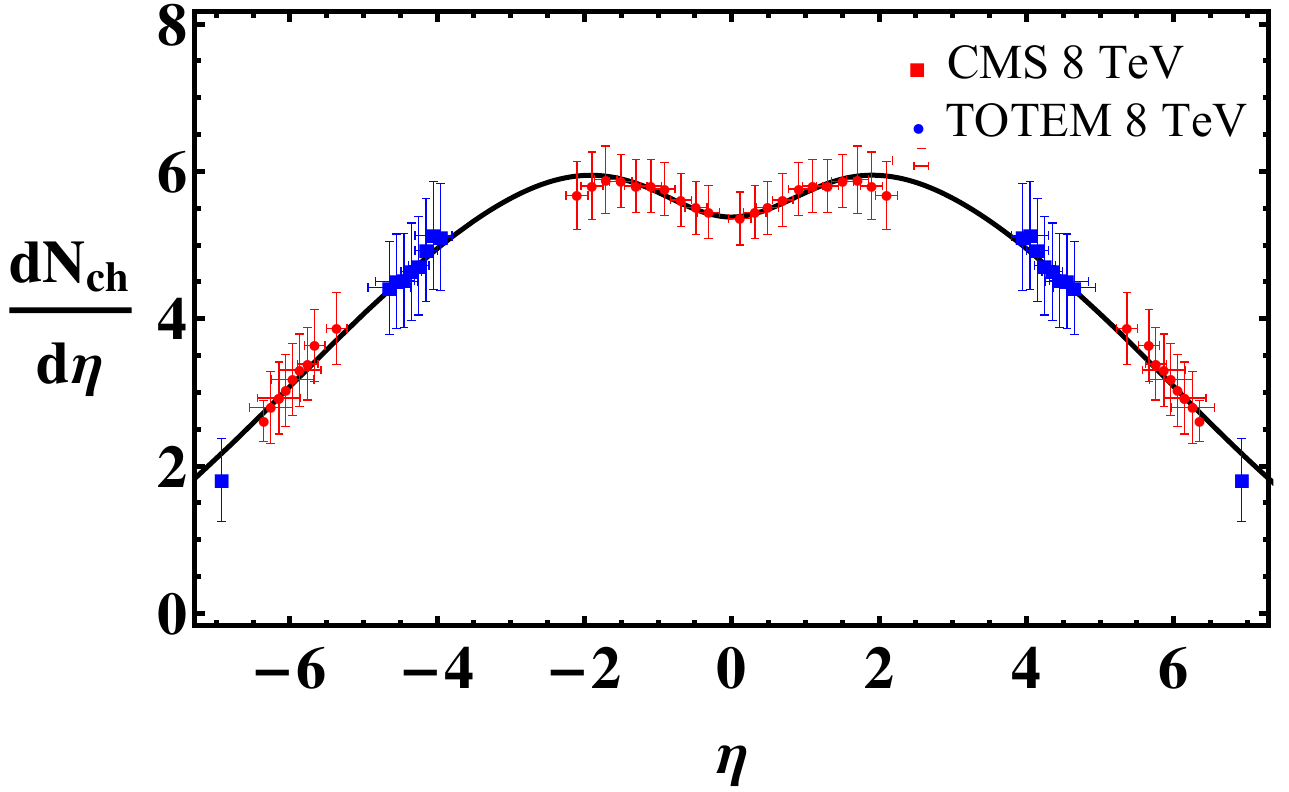}
\caption{\label{f:totemcmsfit}Left: Charged particle $\frac{dN}{d\eta}$ distributions
from CMS~\cite{Khachatryan:2010us} and TOTEM~\cite{Aspell:2012ux} at 7 TeV fitted with
the result of our relativistic hydrodynamical solution described in this paper, with
$\lambda = 1.076$. Right: Charged particle $\frac{dN}{d\eta}$ distributions from CMS
and TOTEM 8 TeV data of Ref.~\cite{Chatrchyan:2014qka,Antchev:2014lez} fitted with 
our results, with $\lambda = 1.066$.}
\end{figure}

Fit to TOTEM $dN/d\eta$ data was shown in Section~\ref{s:totemfit} and in particular in
Fig.~\ref{f:totemfit}. This shows an advanced estimate and reaches out to a large
$\eta$ region. However, it may seem necessary to put more attention on central $\eta$
region and perform fits on a combined central+forward $\eta$ region. Thus in this section
we present $dN/d\eta$ distributions of charged particles from combined TOTEM+CMS datasets.
In the left panel of Fig.~\ref{f:totemcmsfit} a fit to such combined data at
7 TeV~\cite{Khachatryan:2010us,Aspell:2012ux} is shown.

This fit to TOTEM+CMS data~\cite{Khachatryan:2010us,Aspell:2012ux} yields an initial flow
acceleration parameter $\lambda= 1.076$ for 7 TeV pp collisions. As seen from the terms on
the right hand side in Eq.~(\ref{e:conjeps}), the corrected initial energy density depends on
the acceleration parameter $\lambda$, the driving force for the hydrodynamic expansion or the
pressure gradients and volume element expansion. The obtained correction factor is 1.273,
thus the advanced hydrodynamic estimate is

\begin{equation}
\epsilon_{\rm corr}(7\;{\rm TeV})=
0.645\;\rm{GeV/fm}^3.
\end{equation}
Here we stress that compared to the $\epsilon_{\rm corr}$
estimate from fitting TOTEM only, the difference less than 1\%,
which is quite reasonable. The fit in the right panel of Fig.~\ref{f:totemcmsfit} shows the hydrodynamic
$dN/d\eta$ distribution and charged particle in 8 TeV pp collision measured by TOTEM and
CMS~\cite{Chatrchyan:2014qka} (including TOTEM data measured in collisions with a displaced interaction
vertex~\cite{Antchev:2014lez}). The resulting acceleration parameter is $\lambda=1.066$,
this corresponds to a correction factor of 1.235 and a final result of corrected initial energy 
density form the advance estimate

\begin{equation}
\epsilon_{\rm corr}(8\;{\rm TeV})=
0.641~\;\rm{GeV/fm}^3
\end{equation}
which differs from the previous estimate based on only TOTEM data by less than 0.5\%.

\section{Summary}
We have shown, that based on an accelerating hydro solution and data of the TOTEM and CMS experiments
at CERN LHC, the advanced estimate of the initial energy density yields a value that below the critical
value of 1 GeV/fm$^3$, but is not
inconsistent with a supercritical state in high multiplicity 7 and 8 TeV
proton-proton collisions. The energy density is proportional to the measured multiplicity, and so in 
high-multiplicity events, initial energy densities several times the
critical energy density of 1 GeV/fm$^3$ have been reached.
It means, that an important and necessary condition is satisfied for the
formation of a non-hadronic medium in high multiplicity ($dN/d\eta_{\eta=0}>9$)
7 and 8 TeV p+p collisions at CERN LHC,
however, the exploration of additional signatures (radial and elliptic
flow, volume or mean multiplicity dependence of the signatures of the
nearly perfect fluid in p+p collisions, scaling of the HBT radii with
transverse mass, and  possible direct photon signal and low-mass dilepton enhancement)
and their multiplicity dependence
can be a subject of detailed experimental investigation even in p+p
collisions at the LHC. It is also important to note, that even the measurement of 
differential elastic scattering cross-section has implications regarding this estimate.

The main result of our study indicates, that the initial energy density is apparently large
enough in high multiplicity p+p collisions at the $\sqrt{s}$ = 7 and 8 TeV
LHC energies to create a strongly interacting quark-gluon plasma, so a
transition with increasing multiplicity is expected, as far as
hydrodynamical phenomena are considered. Since in RHIC p+p collisions,
multiplicities are a factor of almost 10 smaller~\cite{Adare:2011vy}, high
enough initial energy densities are most probably not reached there. However,
that does not necessarily mean that no collectivity is found in those systems.
As for p+A collisions and similar, intermediate size systems, the situation is
more close to nucleus-nucleus collisions, as indicated by Ref.~\cite{Adare:2015bua}.

Probably the most important implication of our study is the need for an e+p
and e+A collider: as far as we know only in lepton induced proton and heavy
ion reactions can one be certain that a hydrodynamically evolving medium
is not created even at the TeV energy range. The results of lepton-hadron
and lepton-nucleus interactions thus will define very clearly the particle
physics background to possible collective effects. For example, recently
azimuthal correlations were observed in high multiplicity p+p and p+A
as well as in heavy ion reactions (the ridge effect~\cite{CMS:2012qk,Padula:2011yk}), whose origin is
currently not entirely clear. If such a ridge effect appears also in e+p
and e+A collisions, then most likely this effect is not of a hydrodynamical
origin, while if it does not appear in e+p and e+A collisions in the same
multiplicity range as in p+p and p+A reactions, than the ridge is more
likely a hydrodynamical effect.

If indeed a strongly interacting non-hadronic medium is formed in high multiplicity p+p collisions,
than purely the jet suppression in heavy ion collisions does not reveal the true
nature of these systems: the proper measure would be energy loss per unit length
(as proposed in Ref.~\cite{Csorgo:2009wc}),
which may be quite similar in these systems, even if the total suppression is different.

We are looking forward to measurements unveiling the nature of
the matter created in proton-proton collisions. In experimental p+p data, one should look for
the enhancement of the photon to pion ratio in high multiplicity events (as compared to low multiplicity
ones)~\cite{Tannenbaum:2010yx}, for a hydrodynamic scaling of Bose-Einstein correlation radii or that of
azimuthal asymmetry~\cite{Adare:2006ti}, or even the enhancement of low mass
dileptons~\cite{Afanasiev:2007aa}.

\vspace{6pt} 

\section*{Acknowledgments}
The authors thank the important discussions to Simone Giani, Paolo Guibellino,
Federico Antinori, Michael Tannenbaum, P\'eter L\'evai, Sandra S. Padula, Endel Lippmaa
and Renato Campanini. We acknowledge the support of the Hungarian OTKA grant NK-101438 and of the
bilateral Chinese-Hungarian governmental project, TeT 12CN-1-2012-0016. T. Cs. gratefully acknowledges
partial support from the Ch. Simonyi Foundation. M. Cs. was supported by the J\'anos Bolyai Research
Scholarship of the Hungarian Academy of Sciences.

\providecommand{\href}[2]{#2}\begingroup\raggedright

\end{document}